\documentclass[preprintnumbers, showpacs, floatfix,onecolumn,preprintnumbers, letterpaper, superscriptaddress,nofootinbib]{revtex4}
\usepackage{amsfonts}
\usepackage{amsmath}
\usepackage{amssymb}
\usepackage{latexsym}
\usepackage{graphicx}
\usepackage{float}
\usepackage{subfigure}
\usepackage{hyperref} 

\begin{document}

\title{Holographic paramagnetic-ferromagnetic phase transition with Power-Maxwell electrodynamics}
\author{B. Binaei Ghotbabadi$^{1}$,  A. Sheykhi $^{1,2}$\footnote{
		asheykhi@shirazu.ac.ir},G. H. Bordbar$^{1}$\footnote{
	ghbordbar@shirazu.ac.ir}}
\address{$^1$ Physics Department and Biruni Observatory, College of
	Sciences, Shiraz University, Shiraz 71454, Iran\\
	$^2$Research Institute for Astronomy and Astrophysics of Maragha
	(RIAAM), P.O. Box 55134-441, Maragha, Iran}
\begin{abstract}
We explore the effects of Power-Maxwell nonlinear electrodynamics
on the properties of holographic paramagnetic-ferromagnetic phase transition in the background of
Schwarzchild Anti-de Sitter (AdS) black hole. For this purpose, we
introduce a massive $2-$form coupled to the Power-Maxwell field.
We perform the numerical shooting method in the probe limit by
assuming the Power-Maxwell and the $2-$form fields do not back
react on the background geometry. We observe that increasing the
strength of the power parameter causes the formation of magnetic
moment in the black hole background harder and critical
temperature lower. In the absence of external magnetic field
and at the low temperatures, the spontaneous magnetization and the
ferromagnetic phase transition happen. In this case, the critical
exponent for magnetic moment is always $1/2$ which is in agreement
with the result from the mean field theory. In the presence of external
magnetic field, the magnetic susceptibility satisfies the
Cure-Weiss law.
\end{abstract}

\pacs{11.25.Tq, 04.70.Bw, 11.27.+d, 75.10.-b}

\maketitle

\section{Introduction}
The AdS/CFT duality provides a correspondence between a strongly
coupled conformal field theory (CFT) in $d$-dimensions and a
weakly coupled gravity theory in ($d+1$)-dimensional anti-de
Sitter (AdS) spacetime \cite{1,2,3}. Since it is a duality between
two theories with different dimensions, it is commonly called
holography. The idea of holography has been employed in the
condensed matter physics to study the various phenomena such as
superconductivity \cite{hartnoll,5,6,7,8}. For describing the
properties of low temperature superconductors, the BCS theory can
work very well \cite{9,10}. The electronic properties of materials
have been studied using the duality in strongly correlated
systems. Recently, the magnetism also have been attracted the
attentions about the duality application to the condensed matter
physics. There are a few works in investigating the magnetism from
the holographic superconductors point of view
\cite{montull,Donos,albash,m.pujo,iqbal}. An example is the
holographic paramagnetic-ferromagnetic phase transition in a
dyonic Reissner-Nordstrom-AdS black brane which was introduced in
Ref. \cite{dyonic}. This model gives a starting point for
exploration of more complicated magnetic phenomena and quantum
phase transition. It was considered that the magnetic moment could
be realized by a real antisymmetric tensor field which is coupled
to the background gauge field strength in the bulk. It was found
that the spontaneous magnetization happens in the absence of
external magnetic field, and it can be realized as the
paramagnetic-ferromagnetic phase transition. This model was
extended by introducing two antisymmetric tensor fields which
correspond with two magnetic sublattices in the materials
\cite{p.Acai6}. In the framework of usual Maxwell electrodynamics,
holographic paramagnetism-ferromagnetism phase transition have
been investigated
\cite{p.Acai6,Coexistence.Cai,Yokoi,Cai3,Cai4,Insulator.Cai,Understanding.Cai,Lifshitz5}.
However, it is interesting to investigate the effects of nonlinear
electrodynamics on the properties of the holographic
paramagnetic-ferromagnetic phase transition. Considering three
types of nonlinear electrodynamics, namely, Born-Infeld,
logarithmic and exponential nonlinear electrodynamics, and using
the numerical methods, it has been observed that in the
Schwarzschild AdS black hole background, the higher nonlinear
electrodynamics corrections make the magnetic moment harder to
form in the absence of external magnetic field \cite{Zhang2,Wu1}.
Although, the properties of holographic superconductor with
conformally invariant Power-Maxwell electrodynamics have been
studied in \cite{PM1,PM2,Shey1,Shey2,Shey3}, the properties of
holographic paramagnetic-ferromagnetic phase transition coupled to
the Power-Maxwell field have not been explored yet.
For applications of gauge/gravity duality like
holographic superconductor model, the gravitational model could
not be studied as well as others which satisfy the behavior of
boundary theory and the condition of string theory. Since in this
viewpoint, the Power-Maxwell nonlinear electrodynamics can be
representation the fields with higher order terms, so it can be
useful for our investigation. In this paper, we are going to
extend the study on the holographic paramagnetic-ferromagnetic
phase transition by taking into account the nonlinear
Power-Maxwell electrodynamics. In particular, we shall investigate
how the Power-Maxwell electrodynamics influence the critical
temperature and magnetic moment. Interestingly, we find that the
effect of sublinear Power-Maxwell field can lead to the easier
formation of the magnetic moment at higher critical temperature.
We shall focus on 4D and 5D holographic paramagnetic-ferromagnetic
phase transition in probe limit by neglecting the back reaction of
both gauge and the $2-$form fields on the background geometry. We
employ the numerical shooting method to investigate the features
of our holographic model.

This paper is organized as follows. In section \ref{setup}, we
introduce the action and basic field equations in the presence of
Power-Maxwell electrodynamics. In section \ref{numst}, we employ
the shooting method for our numerical calculation and obtain the
critical temperature and magnetic moment. In that section, we also
study the magnetic susceptibility density. In the last section, we
summarize our results.
\section{Holographic Set-up}\label{setup}
We consider a holographic ferromagnetism model in Einstein gravity
in a $d$-dimensional AdS spacetime which is given by the action,
\begin{eqnarray}
I&=&\frac{1}{2\kappa ^{2}}\int
d^{d}x\sqrt{-g}\left(R-{2}{\Lambda}+ L_{1}\left( F\right)
+\lambda^{2} L_{2}\right) ,  \notag \\
&&\label{Act}
\end{eqnarray}%
where $\kappa ^{2}=8\pi G$ with $G$ is Newtonian gravitational
constant, $g$ is the determinant of metric, $R$ is Ricci scalar
and $\Lambda=-{(d-1)(d-2)}/{2l^{2}}$ is the cosmological constant
of $d-$dimensional AdS spacetime with radius $l$.
$L_{1}(F)=(-F)^{\alpha}/4$, where $F=F_{\mu \nu }F^{\mu \nu }$ in
which $F_{\mu \nu }=\nabla _{\lbrack \mu }A_{\mu ]}$ and $A_{\mu}$
is the gauge potential of U(1) gauge field and $\alpha$ is the
power parameter of the Power-Maxwell field. In the case where
$\alpha$ tend to zero the Power-Maxwell Lagrangian will reduce to
the Maxwell case ($L_{1} \rightarrow -F_{\mu\nu}F^{\mu\nu}/4$) and
the Einstein-Maxwell theory is recovered. Besides,
the Power-Maxwell action is invariant under conformal
transformation $g_{\mu\nu}\to\Omega^{2}g_{\mu\nu}$ and $A_{\mu}\to
A_{\mu}$ in $d$-dimension is given by
\begin{equation}
I_{M}=\frac{1}{8\kappa ^{2}}\int{d^{d}x \sqrt{-g}(-F)^\alpha},
\end{equation}
where $F$ is the Maxwell invariant. The
associated energy-momentum tensor of the above action is given by
\begin{equation}
T_{\mu\nu}=2(\alpha
F_{\mu\rho}F{^{\rho}_{\nu}}F^{\alpha-1}-\frac{1}{4}g_{\mu\nu}F^{\alpha}).
\end{equation}
One can easily check that the above energy-momentum tensor is
traceless for $\alpha={d}/{4}$. In this paper, for generality, we
consider not only the conformal case, but also the arbitrary value
of $\alpha$. This allows us to consider more solutions from
different aspects \cite{shamsip30,dehyadegari}. The term $L_{2}$
in action (\ref{Act}) is defined as \cite{Cai3}
\begin{eqnarray}{\nonumber}
L_{2}=-\frac{1}{12}(dM)^{2}-\frac{m^{2}}{4}M_{\mu\nu}M^{\mu\nu}-\frac{1}{2}M^{\mu\nu}F_{\mu\nu}
-\frac{J}{8}V(M),
  \label{Act1}
\end{eqnarray}
where $\lambda$ and $J$ are two constants with $J<0$ for producing
the spontaneous magnetization and $\lambda^{2}$ characterizes the
back reaction of the two polarization field $M_{\mu\nu}$ and the
Maxwell field strength on the background geometry. In addition,
$m$ is the mass of $2$-form field $M_{\mu\nu}$ being greater than
zero \cite{Cai3} and $dM$ is the exterior differential $2$-form
field $M_{\mu\nu}$. The nonlinear potential of $2$-form field
$M_{\mu\nu}$, $V(M_{\mu\nu})$, describes the self interaction of
polarization tensor which should be expanded as the even power of
$M_{\mu\nu}$. In this model, we take the following form for the
potential
 \begin{equation}
 V\left(M\right) =\left( ^{*}M_{\mu\nu}{M^{\mu\nu}}\right)^{2}=[^{*}(M \wedge M)]^{2},
 \label{potential}
\end{equation}%
where $*$ is the Hodge star operator. We choose this form just for
simplicity. This potential shows a global minimum at some nonzero
value of $\rho$ \cite{Cai3}.

Varying the action (\ref{Act}) with respect to $M_{\mu\nu}$ and
$A_{\nu}$, the field equation read, respectively,
\begin{eqnarray}
 0&=& \nabla ^{\tau }(dM)_{\tau\mu\nu}-\notag \\
 &&-m^{2}M_{\mu\nu}-J (^{*}M_{\tau\sigma}M^{\tau\sigma})(^{*}M_{\mu\nu})-F_{\mu\nu} \,  \label{01} \\
 0&=& \nabla ^{\mu }\left(\alpha{F_{\mu\nu}}{(F)^{\alpha-1}}+\frac{\lambda^{2}}{4}M_{\mu\nu}\right). \,  \label{02}
\end{eqnarray}
In the probe limit, we can neglect the back reaction of the 2-form
field. As the background geometry, we consider the $d-$dimensional
Schwarzschild AdS black hole which its metric reads
\begin{equation}
ds^{2}=L^{2}\left(-r^{2}f(r)dt^{2}+\frac{dr^{2}}{r^{2}f(r)}+r^{2}\sum_{i=1}^{d}dx_{i}^{2}\right),
\end{equation}%
with
\begin{equation}
 f(r)=1-\left(\frac{r_{+}}{r}\right)^{d-1},
\end{equation}%
where $r_{+}$ is the event horizon radius of the black hole. The
Hawking temperature of black hole on the horizon which will be
interpreted as the temperature of CFT, is given by \cite{q.pan}
\begin{equation}
T=\frac{f^{\prime }(r_{+})}{4\pi }=\frac{(d-1)r_{+}}{4\pi },
\label{3}
\end{equation}%
In order to explore the effects of the power parameter $\alpha$ on
the holographic ferromagnetic phase transition, we take the
self-consistent ansatz with matter fields as follows,
\begin{equation}
M_{\mu\nu}=-p(r)dt{\wedge}dr+ \rho(r)dx{\wedge}dy ,  \label{M}
\end{equation}%
\begin{equation}
A_{\mu}=\phi(r)dt+ B xdy,  \label{A}
\end{equation}%
where $B$ is a constant magnetic field which is considered as an
external magnetic field of dual boundary field theory. Inserting
this ansatz into Eqs. (\ref{01}) and (\ref{02}), we arrive at
\begin{eqnarray}
0 &=&\rho ^{\prime \prime }+\rho ^{\prime }\left[ \frac{%
f^{\prime }}{f}+\frac{d-2}{r}\right] -\frac{\rho
}{r^{2}f}\left[m^{2}+4Jp^{2}\right]+\frac{B}{r^{2}f} ,
    \notag \\
0 &=&\left(m^{2}-\frac{4J\rho^{2}}{r^4}\right)p-\phi ^{\prime}, \notag \\
0&=& \phi ^{\prime \prime}+\frac{2 \phi^{\prime}}{r^{9}}\left[\frac{\frac{d-2}{2}\phi^{\prime 2}-
(2\alpha-\frac{d+2}{2})\frac{B^{2}}{r^{4}}}{(2\alpha-1)\phi^{\prime 2}-\frac{B^{2}}{r^{4}}}\right]+\notag\\
&&+\frac{\lambda^{2}}{2^{\alpha+1}\alpha
}\left(p^{\prime}+\frac{2}{r}p\right)\left[\frac{(\phi^{\prime
2}-\frac{B^{2}}{r^{4}})^{2-\alpha}}{(2\alpha-1)\phi^{\prime
2}-\frac{B^{2}}{r^{4}}}\right]
 , \,\label{EOM}
  \end{eqnarray}%
where the prime denotes the derivative with respect to $r$.
Obviously, the above equations reduce to the corresponding
equations in Ref.~\cite{Cai3} when $\alpha\rightarrow 1$ and
$d=4$. We should specify boundary conditions for the fields to
solve Eq. (\ref{EOM}) numerically. At the horizon, we need to
impose a regular boundary condition. Therefore, in additional to
$f(r_{+})=0$, because the norm of the gauge field, namely
$g_{\mu\nu}A^{\mu}A^{\nu}$,  should be finite at the horizon, we
require $\phi(r_{+})=0$ and
$\rho(r_{+})=\frac{(d-1)r_{+}}{m^{2}}\rho^{\prime}+\frac{B}{m^{2}}$.
The behaviors of model functions governed by the field
equations (\ref{EOM}) near the boundary ($r\rightarrow \infty $)
are given by
\begin{gather}
\phi (r)\sim \mu
-\frac{\sigma^{\frac{1}{2\alpha-1}}}{r^{\frac{(d-2)}{2\alpha-1}-1}},\qquad
p(r)\sim \frac{\sigma}{m^{2}r^{(d-2)}},\notag\\
\rho (r)\sim \frac{\rho _{-}}{%
r^{\Delta _{-}}}+\frac{\rho _{+}}{r^{\Delta
_{+}}}+\frac{B}{m^{2}}, \label{boundval}
\end{gather}%
where $\mu $ and $\sigma $ are respectively interpreted as the
chemical potential and charge density of dual field theory, and
\begin{equation}
\Delta _{\pm
}=\frac{1}{2}\left[(5-d)\pm\sqrt{4m^{2}+16+(d-1)(d-9)}\right].
\end{equation}
According to AdS/CFT correspondence, $\rho_{+}$ and $\rho_{-}$ are
two constants correspond to the source and vacuum expectation
value of dual operator when $B=0$. Therefore, condensation happens
spontaneously below a critical temperature when we set
$\rho_{+}=0$. By considering $B\ne0$, the asymptotic behavior is
governed by external magnetic field $B$. It is important to note
that the boundary condition for the gauge field $\phi$ depends on
the power parameter $\alpha$ of the Power-Maxwell field unlike
other nonlinear electrodynamics such as Born-Infeld-like
electrodynamics \cite{Zhao, Jing}. Using boundary condition Eq.
(\ref{boundval}) and the fact that $\phi$ should be finite as
$r\rightarrow \infty$, we require that
$\frac{d-2}{2\alpha-1}-1>0$, which restricts the values of
$\alpha$ to be $\alpha<\frac{d-1}{2}$. On the other hand since
$\frac{d-2}{2\alpha-1}>1$, it must be a positive real number, it
leads to the range of the parameter $\alpha$ to be
$1/2<\alpha<\frac{d-1}{2}$.

{ In order to see the main properties of the model, we consider
the probe limit for simplicity. It is worth noting that we have
two kinds of probe limit. In the former case, one may take the
model parameter $\lambda\rightarrow0$ as in Ref. \cite{dyonic} by
neglecting the back-reaction of the massive $2$-form field on the
background geometry and the Maxwell field. In that case, the
effect of the Maxwell field on the background geometry has been
considered. In this probe limit, the influence of external field
on the materials is considered, but they neglect the back reaction
of the electromagnetic response on the external field and
structures of materials. In the latter case one may neglect all
back reaction of matter fields including the Maxwell field on the
background geometry. In our model, we follow the latter type of
the probe limit in which the interaction between the
electrodynamic response and external field is taken into account
and the $\lambda$ parameter can take any small value.
The charge density of the system is given by
\cite{Cai4}
\begin{equation*}
\sigma=-\frac{\lambda^{2} \mu}{4 m^{2}}+\alpha \mu \left(\frac{2
B^2}{r^{4}}-\frac{2 \mu{^2}}{r^{4}}\right)^{\alpha-1}.
\end{equation*}
We see that the properties of this black hole solution depends on
the value of $1-\frac{\lambda^{2}}{4m^{2}}$ when $B\to 0$ and $\alpha\to 1$ in above equation.
 To investigate
the physical properties when $1-\frac{\lambda^{2}}{4m^{2}}\neq 0$,
one can compute the partial derivative of the charge density with
respect to chemical potential and it is easy to check that when
$1-\frac{\lambda^{2}}{4m^{2}}>0$ the value of this partial
derivative become positive \cite{Cai4}. It
gives a chemical stable dual boundary system. For $1-\frac{\lambda^{2}}{4m^{2}}<0$, this parameter is
negative which leads the dual boundary system to be chemical
instability. So in this probe limit, the parameters have to
satisfy the condition $\lambda^{2}<4m^{2}$. As an
example, we choose $\lambda=1/2$ which is small enough for this
parameter.

In the following sections, we will study the holographic
ferromagnetic-paramagnetic phase transition numerically.
  \begin{figure*}[p]
\centering{%
\subfigure[~$\protect$D=4$$]{
 \label{fig1a}\includegraphics[width=.36\textwidth]{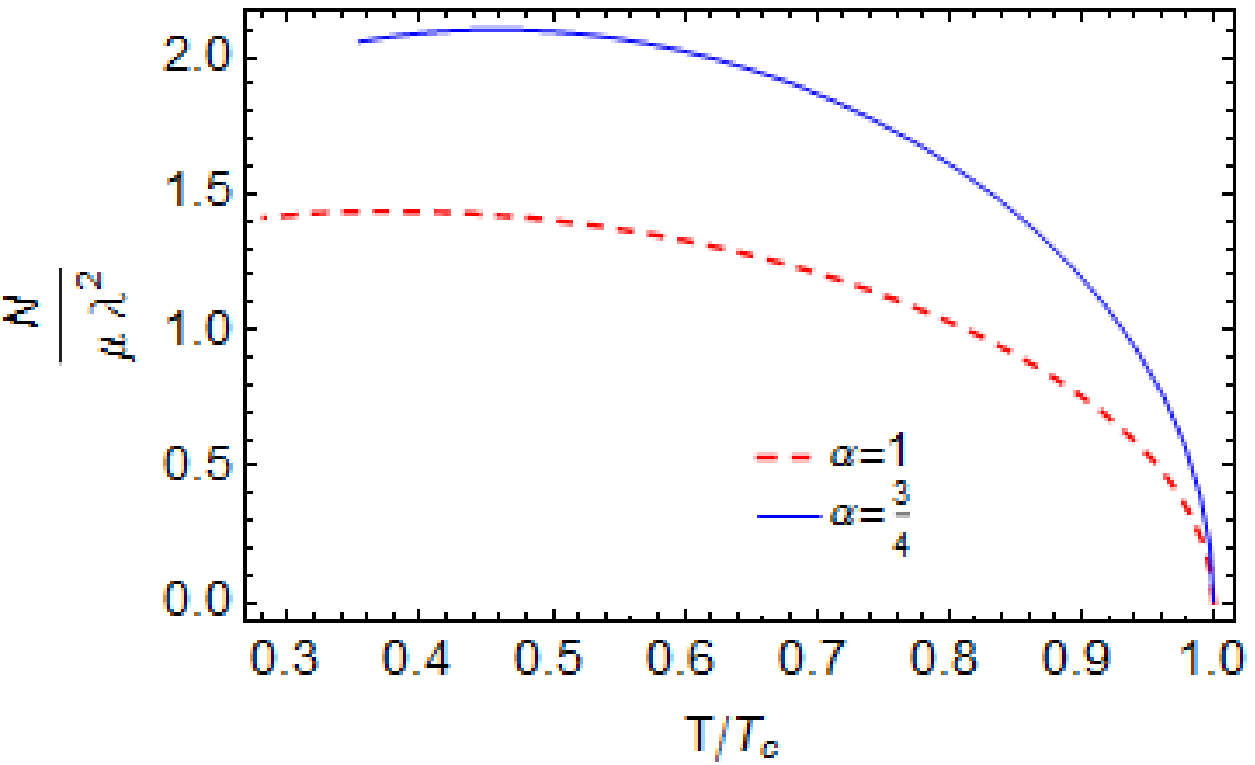}\qquad}%
 \subfigure[~$\protect$D=5$$]{
 \label{fig1b}\includegraphics[width=.36\textwidth]{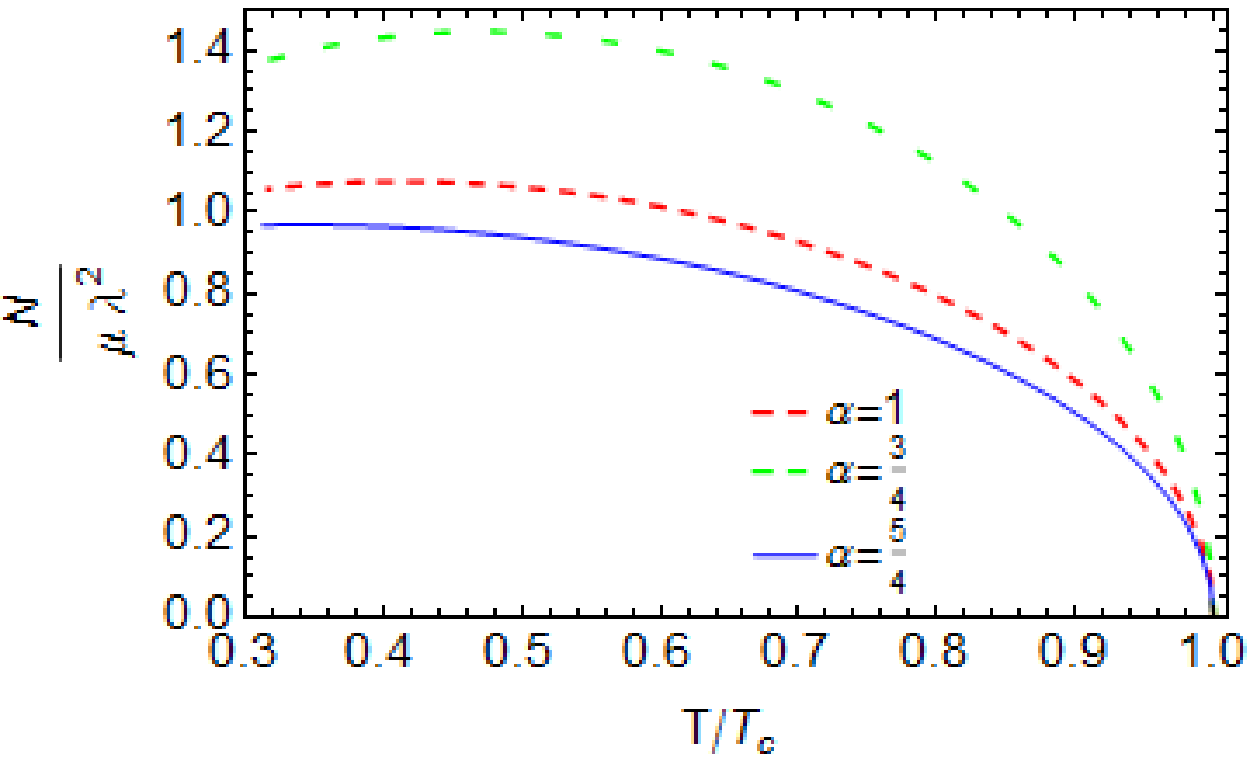}\qquad}
 \subfigure[~$\protect$D=4$$]{
 \label{fig1c}\includegraphics[width=.36\textwidth]{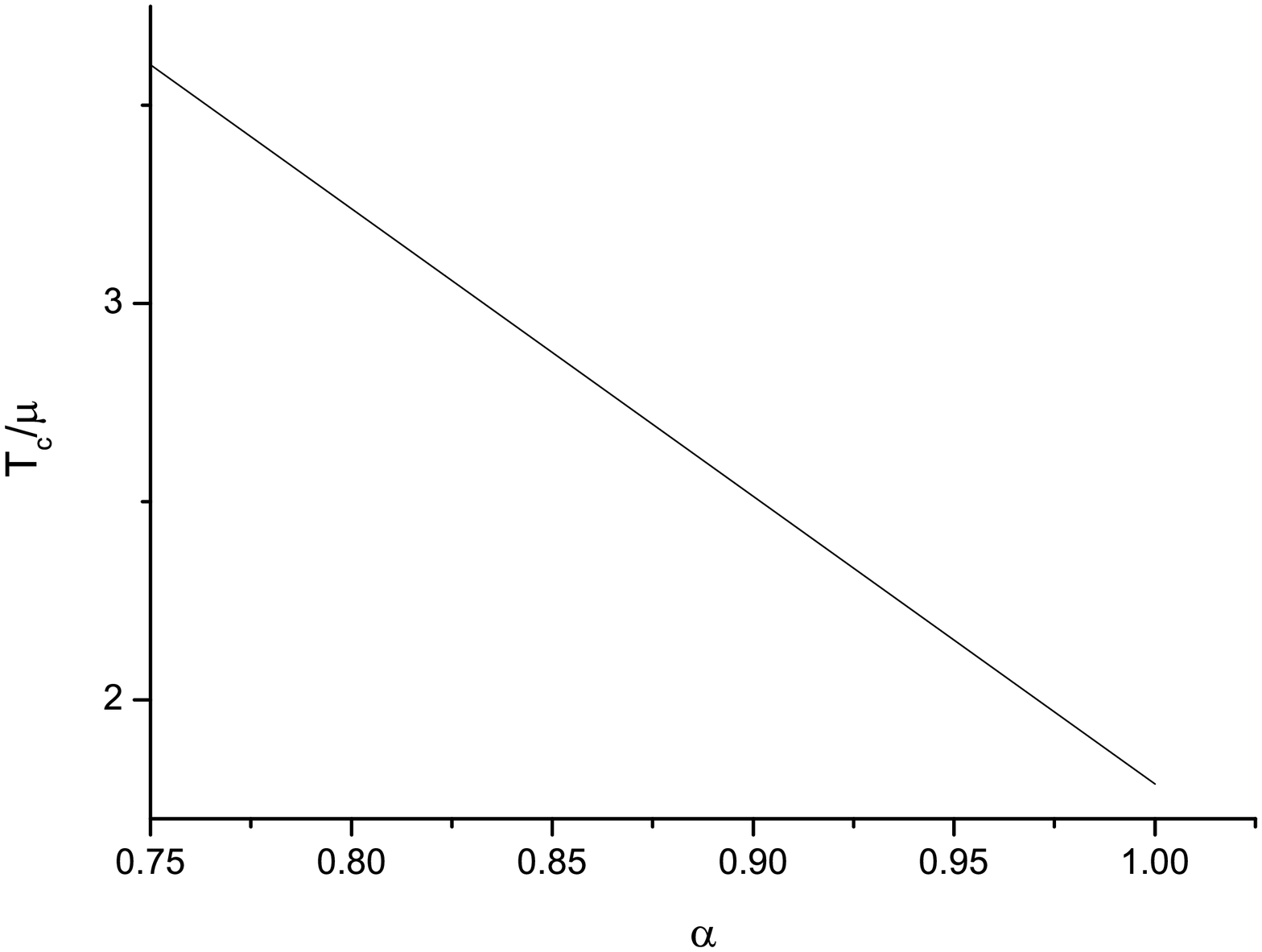}\qquad}%
 \subfigure[~$\protect$D=5$$]{
 \label{fig1d}\includegraphics[width=.36\textwidth]{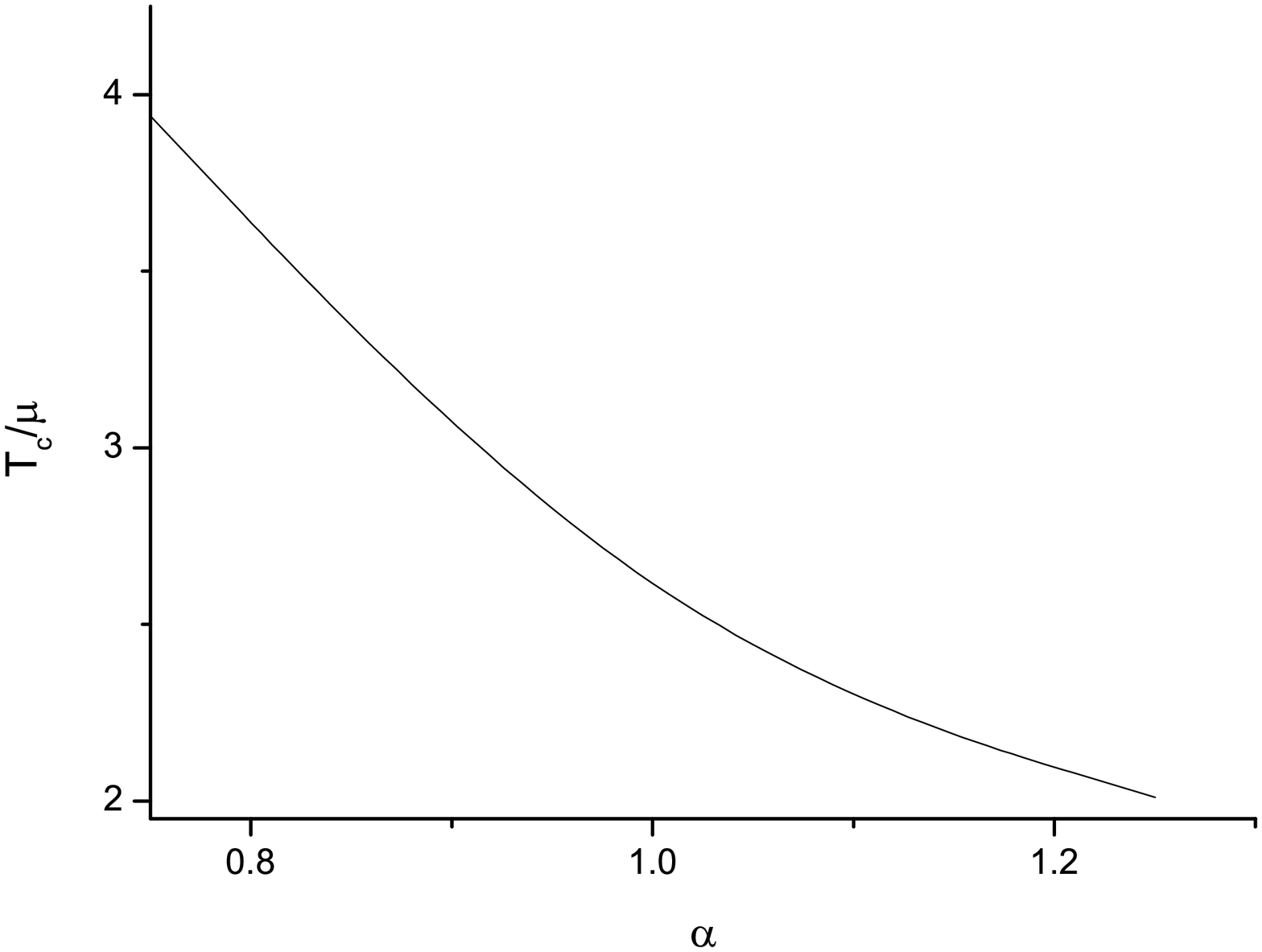}\qquad}}
\caption{The behavior of magnetic moment $N$ and the critical
temperature in two dimensions with different values of $\alpha$
for $m^{2}=1/8$ and $J=-1/8$.}
\label{fig1}
\end{figure*}
\begin{table*}[p]
\centering%

\begin{tabular}{llllll}
\hline
$\alpha$ &  $ 3/4 $& $ $ & $1 $  & $ $ &$5/4$ \\
\hline
 $4D $  &  $ 3.6023 $& $ $ & $1.7870$ & $ $  & $-$   \\
 $ 5D$ &  $ 3.9388 $& $ $  & $2.4367$  & $ $ & $2.0103$   \\
 \hline

 \end{tabular}

 \label{Table1}
  \caption{Numerical results of ${T_{c}}/{\protect\mu }$ for
different values of $\protect\alpha $ in 4D and 5D.}
 \end{table*}

 \begin{table*}[p]
 \centering%
 \begin{tabular}{llllll}
 \hline
 $\alpha$&$ 3/4 $ & $1 $ & $5/4$   \\
 \hline
 $4D$& $ 6.0253(1-T/Tc)^{1/2} $ & $2.9409(1-T/Tc)^{1/2}$  & $-$   \\
 $5D$& $ 3.8213(1-T/Tc)^{1/2} $ & $2.9810(1-T/Tc)^{1/2}$  & $2.6211(1-T/Tc)^{1/2}$   \\
 \hline
 \end{tabular}
 \label{Table2}
 \caption{The magnetic moment $N$ with different values of $\alpha$ in 4D and 5D.}
 \end{table*}

 \begin{figure*}[t]
 \centering{%
 \subfigure[~$\protect$$D=4,$$ \alpha=3/4$]{
 \label{fig2a}\includegraphics[width=.36\textwidth]{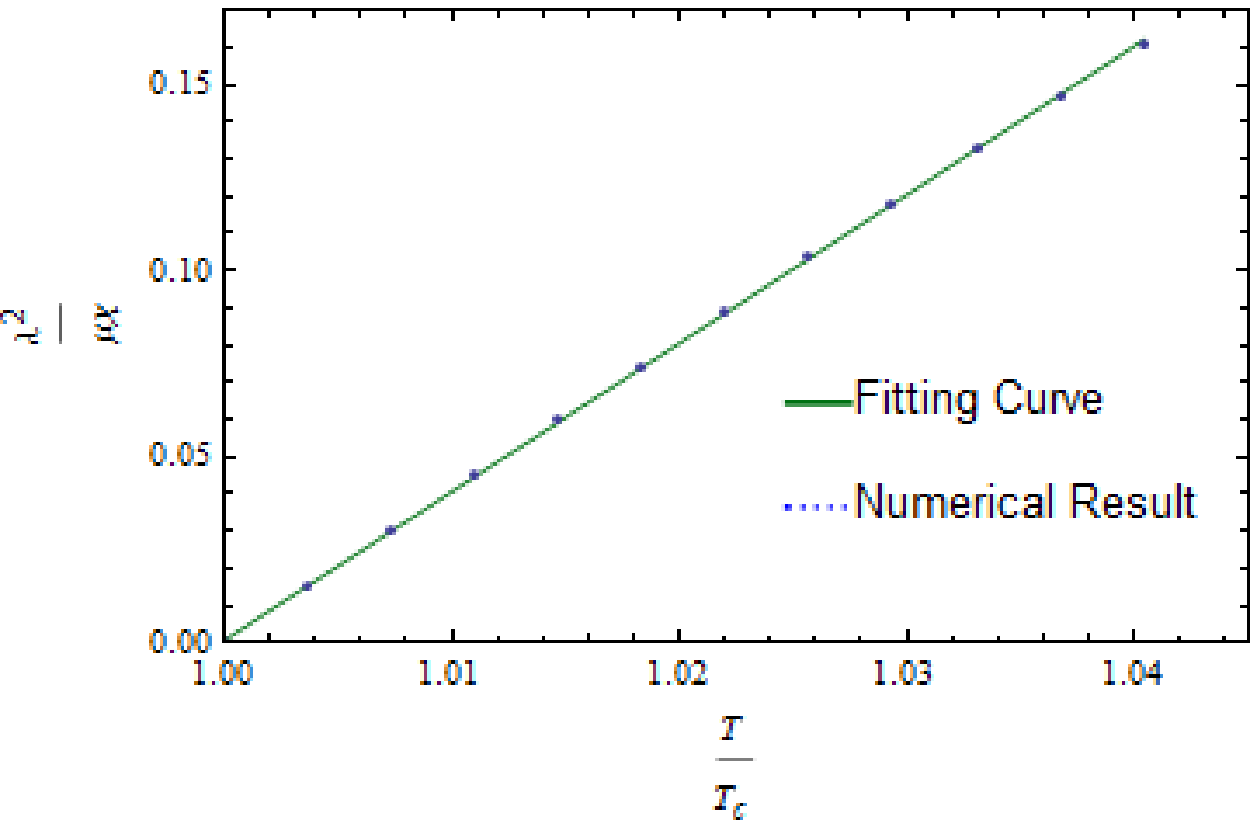}\qquad}
 \subfigure[~$\protect$$D=4$, $ \alpha=1$]{
 \label{fig2b}\includegraphics[width=.36\textwidth]{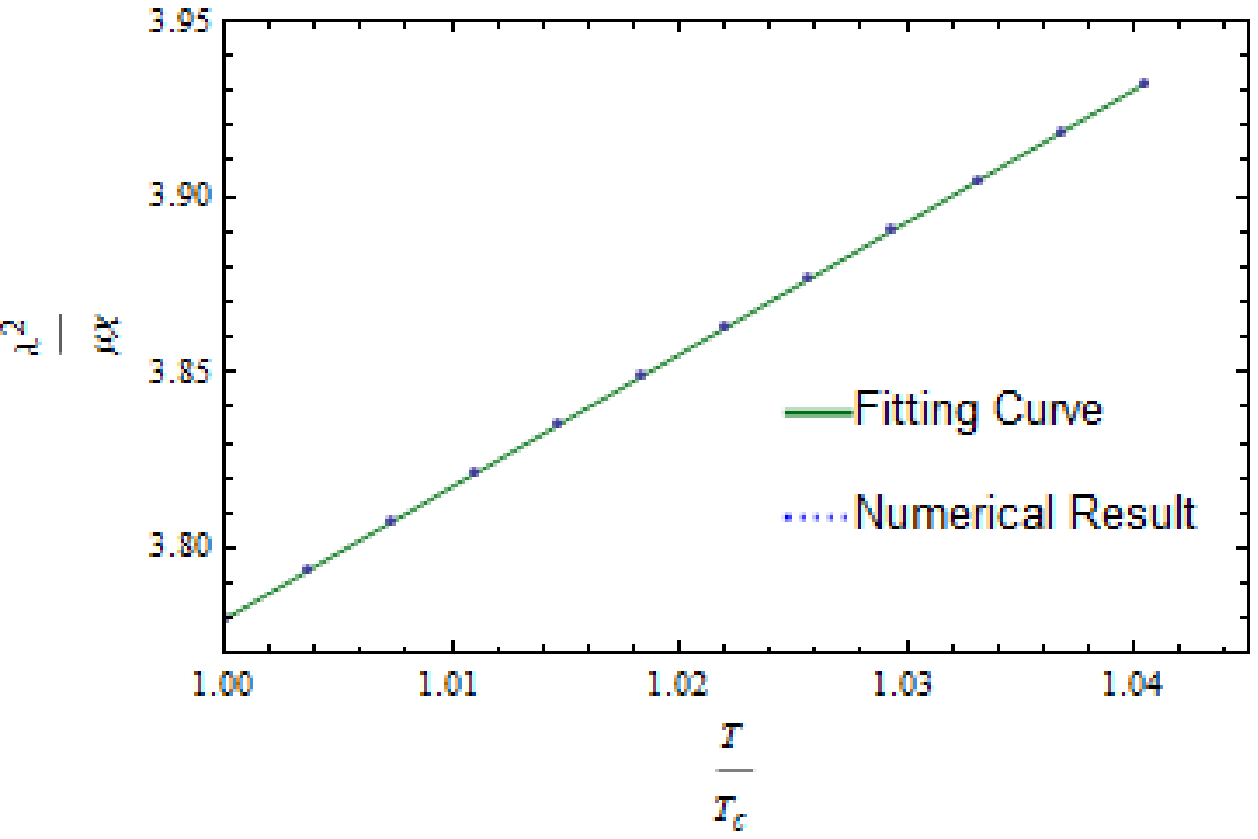}\qquad}
 \subfigure[~$\protect$$D=5,$$ \alpha=3/4$]{
 \label{fig2c}\includegraphics[width=.36\textwidth]{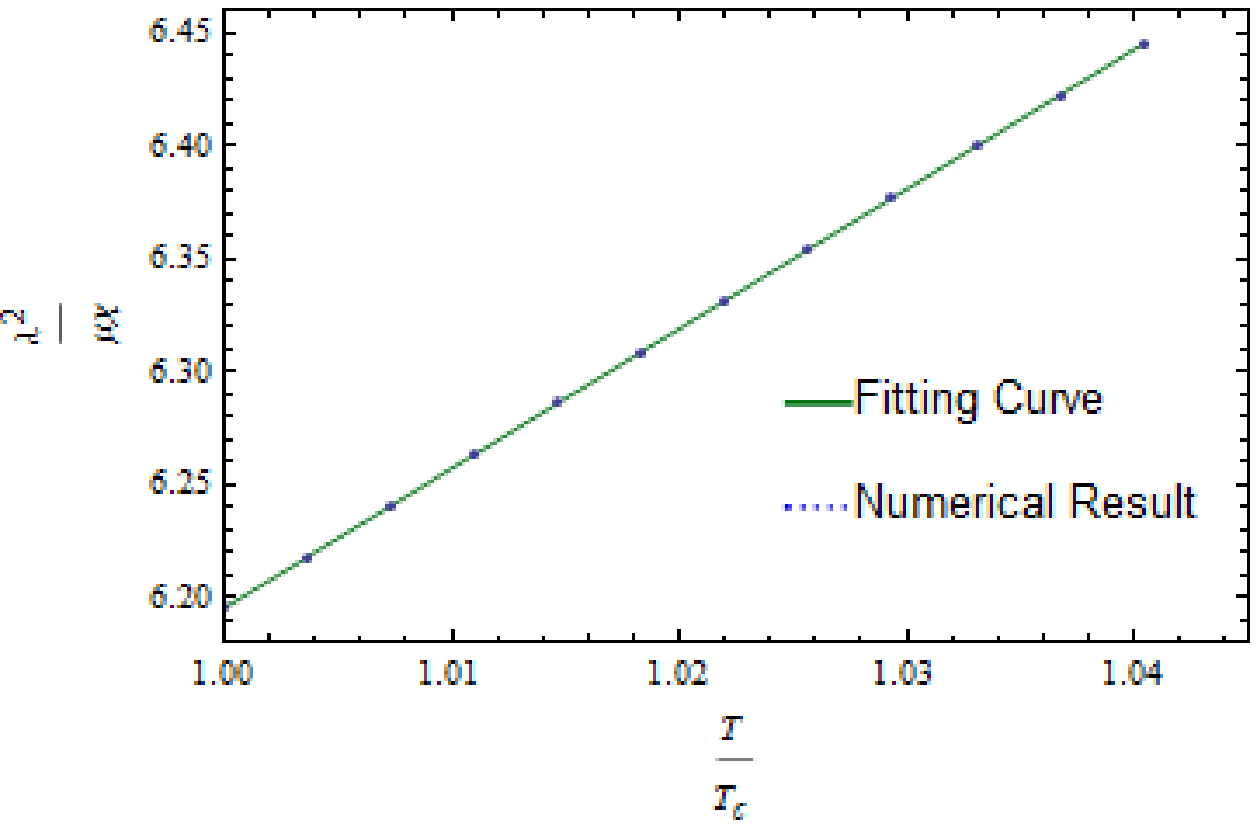}\qquad}
 \subfigure[~$\protect$$D=5$, $ \alpha=1$]{
 \label{fig2d}\includegraphics[width=.36\textwidth]{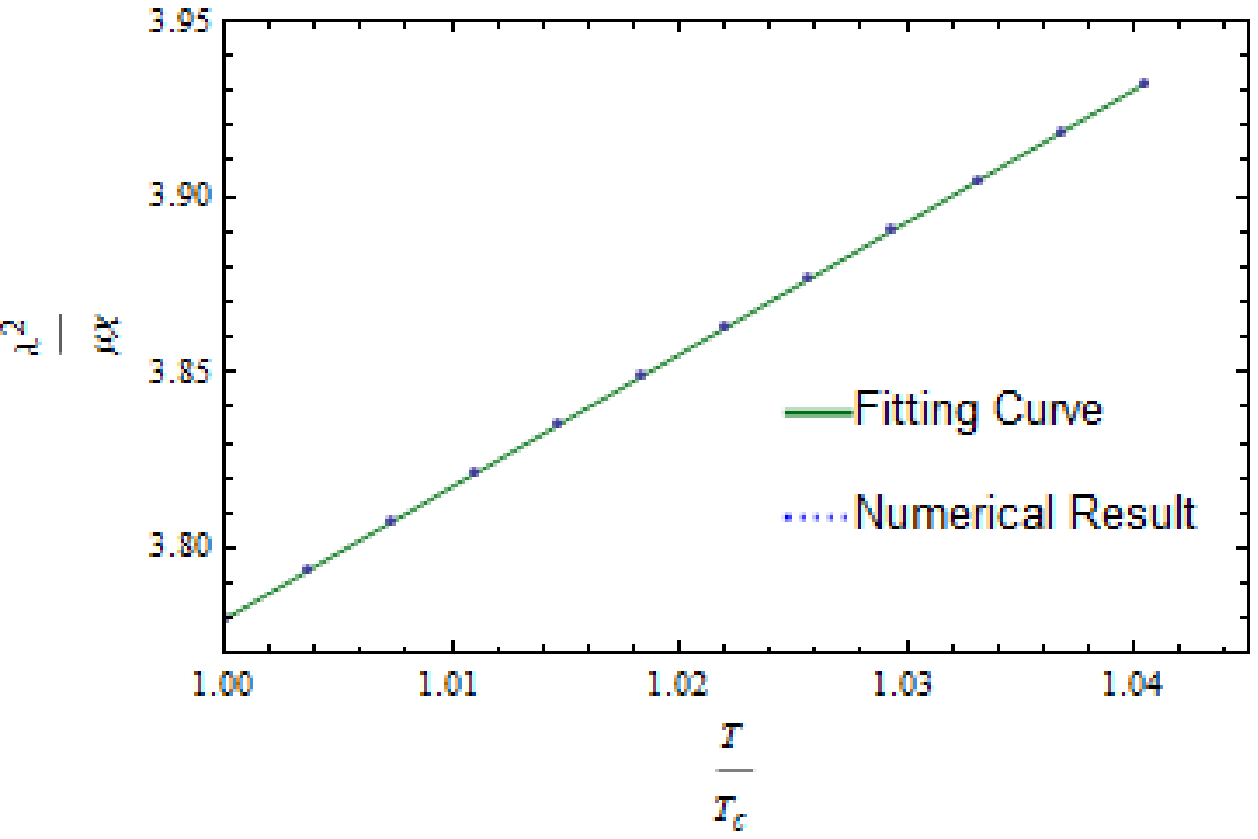}\qquad}
 \subfigure[~$\protect$$D=5,$$ \alpha=5/4$]{
 \label{fig2e}\includegraphics[width=.36\textwidth]{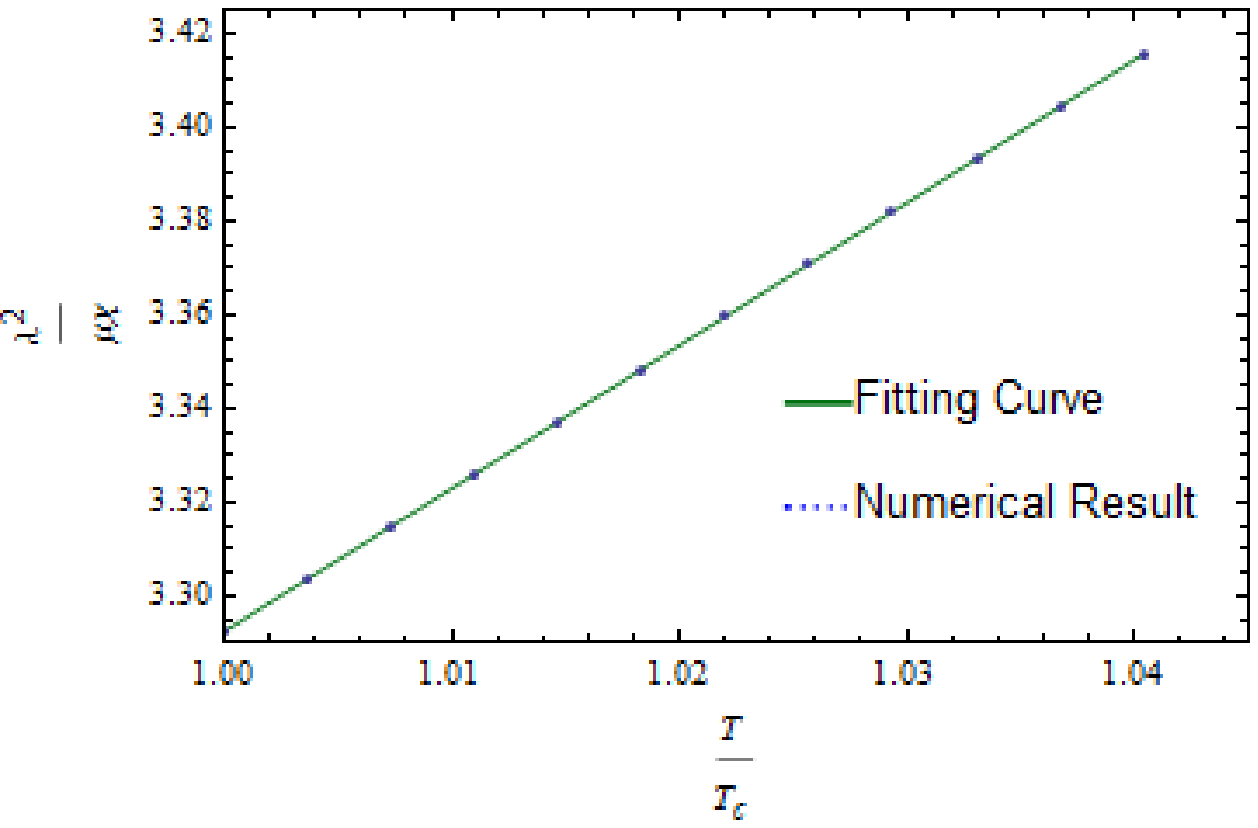}\qquad}}%
\caption{The magnetic susceptibility as a function of temperature
with different values of $\alpha$ in 4D and 5D.} \label{fig2}
\end{figure*}
\section{Numerical calculation for spontaneous magnetization and susceptibility\label{numst}}
In this paper, we work in the grand canonical ensemble where the
chemical potential $\mu$ is a fixed quantity. We have to solve Eq.
(\ref{EOM}) to get the solution of the order parameter $\rho$, and
then compute the value of magnetic moment $N$, which is defined by
\begin{equation}
 N=-\lambda^{2}\int\frac{\rho}{2 r^{d-2}}dr.
 \end{equation}
Since different parameters will give similar results, here we
choose $m^{2}=-J=1/8$ and $\lambda=1/2$ as a typical example in
the numerical computation. In this section, we employ the shooting
method \cite{hartnoll} to numerically investigate the holographic
phase transition. Hereafter, we define the dimensionless
coordinate $z=r_{+}/r$ instead of $r$, since it is easier to work
with it. In terms of this new coordinate, $z=0$ and $z=1$
correspond to the boundary and horizon respectively. Besides
setting $l$ to unity, we also set $r_{+}=1$ in the numerical
calculation, for simplicity, which may be justified by virtue of
the field equation symmetry
 \begin{equation*}
 r\rightarrow ar,\text{ \ \ \ \ }f\rightarrow f,\text{ \ \ \ \ }\phi
 \rightarrow a\phi ,
 \end{equation*}%
First, we expand Eqs.(\ref{EOM}) near black hole horizon ($z=1$)%
\begin{eqnarray}
\rho &\approx &\rho (1)+\rho ^{\prime }(1)(1-z)+\frac{\rho ^{{\prime }{%
\prime }}}{2}(1-z)^{2}+\cdots , \\
\phi &\approx &\phi ^{\prime }(1)(1-z)+\frac{\phi ^{{\prime }{\prime }}}{2}%
(1-z)^{2}+\cdots .
\end{eqnarray}%
In above equations, we have imposed $\phi (1)=0$. In our numerical
process, we will find $\rho (1)$, $\phi ^{\prime}(1)$ such that
the desired values for boundary parameters in Eq. (\ref{boundval})
are attained. At boundary, we set $\rho _{+} $ equal to zero as
the source and consider $\rho _{-}$ as the expectation value of
the magnetic moment. We consider the cases of different power
parameter $\alpha$ in $4D$ spacetime as examples, and then extend
to the case of $5D$ spacetime. We present our results in Fig.
\ref{fig1}. From the left panel of this figure we observe the
magnetic moment with two different values of power parameters in
$d=4$ dimension. Our numerical calculation is presented for $d=5$
dimension in the right panel of Fig. \ref{fig1}. When the
temperature is lower than $T_{C}$, the spontaneous magnetization
appears in the absence of external magnetic field. In the vicinity
of critical temperature, the numerical results show that the
second order phase transition happen which its behavior obtain by
fitting this curve ($N\propto\sqrt{1-T/T_{C}}$). The results have
been shown in Table \ref{Table2}. We find that there is a square
root behavior for the magnetic moment versus temperature, and it
can be found that the critical exponent($1/2$) is the same as the
one from the mean field theory. In other words, the holographic
paramagnetic-ferromagnetic phase transition exists by considering
the Power-Maxwell electrodynamics similar to the cases of
Born-Infeld-like nonlinear electrodynamic discussed in Ref.
~\cite{Zhang2}. As it can be found from this Figure, the magnetic
moment decreases with increasing the power parameter $\alpha$. It
means that the magnetic moment is harder to be formed which is in
a good agreement with similar works  \cite{Zhang2,Wu1}. This
behavior has been seen for the holographic superconductor in the
Schwarzschild- AdS black hole, where the three types of nonlinear
electrodynamics make scalar condensation harder to be formed
\cite{Zhao}.

In Table \ref{Table1}, our numerical results for critical
temperature with different values of power parameter $\alpha$ and
for two cases of dimensions($d=4,5$) are presented. In the Maxwell
limit ($\alpha\rightarrow 1)$, our numerical results reproduce the
ones of \cite{Zhang2} for $d=4$. We see from Table \ref{Table1}
that the critical temperature $T_{c}$ increases by decreasing
the power parameter for fixed dimension. As the power parameter
$\alpha$ becomes larger, the critical temperature decreases. It
means that the magnetic moment is harder to be formed. This
behavior have been reported previously in \cite{Zhang2} too. Fig.
\ref{fig1} confirms above results.

The behavior of susceptibility density of the material in the
external magnetic field is a remarkable characteristic properties
of ferromagnetic material. The static susceptibility density is
defined by
\begin{equation}
 \chi=\lim\limits_{B\to 0}\frac{\partial N}{\partial B}.
\end{equation}
In the presence of magnetic field, the function $\rho$ is nonzero
at any temperature. The magnetic susceptibility obtained by
solving the Eq. (\ref{01}) based on the previous analysis which one
has been discussed in Ref. \cite{Cai3}. We need to shoot for
boundary conditions with one parameter $\rho(r_{+})$ for computing
the susceptibility density. Fig. \ref{fig2} shows the behavior of
susceptibility density near the critical temperature for $4D$ and
$5D$. One can see that when the temperature decreases, $\chi$
increases. In the region of $T\to T^{+}_{c}$, the susceptibility
density satisfies the cure-Weiss law of ferromagnetism
\begin{equation}
\chi=\frac{C}{T+\theta},\text{ \ \ \ \ } T>T_{C}, \theta<0,
\end{equation}
where $C$ and $\theta$ are two constants. The results have been
presented in Table \ref{table3}. Obviously we can seen that the
coefficient in front of $T/T_{c}$ for $1/\chi$ increases, when the
power parameter($\alpha$) decreases.
\begin{table*}[th]
\centering%

\begin{tabular}{llllll}
\hline
& $\alpha$  & $3/4 $ & $1$ & $5/4$  \\
\hline
 $4D$& $ {\lambda^{2}/\chi\mu} $ & $3.7667(T/Tc+0.0034)$  & $3.9906(T/Tc-1)$  & $-$  \\
 $$& $ {\theta/\mu} $ & $0.0124039$  & $-1.7871$  & $-$  \\
 $5D$& $ {\lambda^{2}/\chi\mu} $ & $6.1814(T/Tc+0.0022)$  & $3.7874(T/Tc+0.0222)$  & $3.04777(T/Tc+0.0803)$  \\
 $$& $ {\theta/\mu} $ & $0.0089$  & $0.05421$  & $0.1614$ \\
\hline
\end{tabular}
\label{table3}
\caption{The magnetic susceptibility $\chi$ with different values
of $\alpha$.}
 \end{table*}

\section{Conclusions}
{In this work, we have studied the properties of one-dimensional
holographic paramagnetic-ferromagnetic phase transition in the
presence of Power-Maxwell electrodynamics. We have also
investigated the effects of different dimensions on the system. We
have performed numerical shooting methods for studying our
holographic model. It was shown that the enhancement in power
parameter of electrodynamic model causes the paramagnetic phase
more difficult to be appeared. This result is reflected from our
data. We observed that the increase of the effects of power
parameter makes the lower values for the critical temperature in
our model. Besides, for smaller values of the power parameter, the
gap in the magnetic moment in the absence of magnetic field, is
larger which in turn exhibits that the condensation is formed
harder. {Besides, in a fixed spacetime dimension, with increasing
the Power-Maxwell parameter, the strength of the electromagnetic
field increases, too. Therefore the magnetic moment is decreased
which causes the condensation to be formed easier.}
 We have
also observed that,the behavior of the magnetic moment is always
as $ (1-T/T_{c})^{1/2}$. This is in agreement with the result from
mean field theory. In the presence of external magnetic field,
the inverse magnetic susceptibility near the critical point
behaves as ($\frac{C}{T+\theta}$) for different values of power
parameters in two dimensions, and therefore it satisfies the Cure-
Weiss law. The absolute value of $\theta$ increases by
increasing the power parameter $\alpha$.}
\begin{acknowledgments}
We thank the Research Council of Shiraz
University. The work of AS has been supported financially by
Research Institute for Astronomy and Astrophysics of Maragha
(RIAAM), Iran.
\end{acknowledgments}

\end{document}